\documentclass[preprint,12pt,authoryear]{elsarticle}

\usepackage{amssymb}
\usepackage{amsmath}
\usepackage{multirow}
\usepackage{pifont}
\usepackage{booktabs}
\usepackage{makecell}
\usepackage{subcaption}
\usepackage{tikz}
\usepackage{float}
\usepackage{threeparttable}

\journal{Engineering Applications of Artificial Intelligence}

\begin{document}

\begin{frontmatter}



\title{Adversarial defense based on distribution transfer}

\author[1,2]{Jiahao Chen}
\ead{xaddwell@zju.edu.cn}

\author[1,3]{Diqun Yan\corref{cor1}}
\ead{yandiqun@nbu.edu.cn}

\author[1,3]{Li Dong}
\ead{dongli@nbu.edu.cn}

\cortext[cor1]{Corresponding author.}

\affiliation[1]{organization={Faculty of Electrical Engineering and Computer Science, Ningbo University},%Department and Organization
            city={Ningbo},
            postcode={315211}, 
            state={Zhejiang},
            country={China}}

\affiliation[2]{organization={College of Computer Science and Technology, Zhejiang University},%Department and Organization
            city={Hangzhou},
            postcode={310027}, 
            state={Zhejiang},
            country={China}}

\affiliation[3]{organization={Key Laboratory of Mobile Network Application Technology of Zhejiang Province},%Department and Organization
            city={Ningbo},
            postcode={315211}, 
            state={Zhejiang},
            country={China}}

\begin{abstract}

The presence of adversarial examples poses a significant threat to deep learning models and their applications. Existing defense methods provide certain resilience against adversarial examples, but often suffer from decreased accuracy and generalization performance, making it challenging to achieve a trade-off between robustness and generalization. To address this, our paper interprets the adversarial example problem from the perspective of sample distribution and proposes a defense method based on distribution shift, leveraging the distribution transfer capability of a diffusion model for adversarial defense. The core idea is to exploit the discrepancy between normal and adversarial sample distributions to achieve adversarial defense using a pretrained diffusion model. Specifically, an adversarial sample undergoes a forward diffusion process, moving away from the source distribution, followed by a reverse process guided by the protected model (victim model) output to map it back to the normal distribution. Experimental evaluations on CIFAR10 and ImageNet30 datasets are conducted, comparing with adversarial training and input preprocessing methods. For infinite-norm attacks with 8/255 perturbation, accuracy rates of 78.1\% and 83.5\% are achieved, respectively. For 2-norm attacks with 128/255 perturbation, accuracy rates are 74.3\% and 82.5\%. Additional experiments considering perturbation amplitude, diffusion iterations, and adaptive attacks also validate the effectiveness of the proposed method. Results demonstrate that even when the attacker has knowledge of the defense, the proposed distribution-based method effectively withstands adversarial examples. It fills the gaps of traditional approaches, restoring high-quality original samples and showcasing superior performance in model robustness and generalization.

\end{abstract}

\begin{graphicalabstract}
\end{graphicalabstract}

\begin{highlights}

\item Our paper introduces a novel defense perspective leveraging sample distribution shifts.

\item We propose a defense method using pretrained diffusion models, effectively countering adversarial attacks.

\item Achieving a remarkable balance between model accuracy, robustness, and generalization, surpassing existing defenses.

\item Our method remains effective even when attackers possess knowledge of the defense, showcasing its robustness.

\end{highlights}


\end{frontmatter}


\section{Introduction}

In recent years, deep neural networks have achieved breakthroughs in various fields such as image processing \citep{krizhevsky2012imagenet}, speech recognition, and text analysis \citep{vaswani2017attention}. However, recent research has indicated persistent robustness issues with deep neural networks (DNNs), particularly their vulnerability to adversarial examples \citep{zou2022perturbation}. The existence of adversarial examples significantly hampers the widespread adoption and development of artificial intelligence technologies, especially in fields with heightened security requirements.

Adversarial examples are generated by introducing subtle adversarial perturbations into original clean samples. Although imperceptible to the human eye, these adversarial perturbations can induce misclassification in DNNs. Initially discovered by \citep{szegedy2013intriguing}, adversarial examples in deep learning models were crafted using adversarial attack algorithms such as Limited-memory Broyden-Fletcher-Goldfarb-Shanno (L-BFGS). Subsequently, \citep{fgsm} introduced the simpler Fast Gradient Sign Method (FGSM), generating adversarial examples by performing gradient ascent on samples and providing an explanation for the existence of adversarial examples. This approach utilizes gradient ascent to produce adversarial examples without complex optimization. \citep{moosavi2016deepfool} proposed the DeepFool algorithm to minimize perturbations generated by adversarial examples, finding the nearest adversarial perturbations to the decision hyperplane. For enhanced attack effectiveness. \citep{madry2017towards} introduced the Projected Gradient Descent (PGD) algorithm, optimizing adversarial perturbations through iterative iterations within a constrained perturbation range, significantly boosting the success rate of adversarial attacks.
Meanwhile, \citep{carlini2017cw} optimized perturbations directly as the attack objective, generating adversarial examples through gradient-based optimization. While achieving high attack success rates and making perturbations difficult to detect, this method generated fragile adversarial examples with poor transferability. To improve attack transferability, \citep{dong2018mifgsm} incorporated historical gradient information for gradient correction during perturbation optimization, achieving better black-box transfer attack capability.
Furthermore, to address the robustness of adversarial examples against image transformations, \citep{xie2019difgsm} and \citep{dong2018mifgsm} enhanced FGSM. The former utilized data augmentation for robust adversarial examples generation, while the latter employed convolutions on gradients to produce translation-invariant examples. To mitigate the impact of network depth on gradient estimation, \citep{huang2019enhancing} conducted attacks using shallow-layer features. The prosperity of various adversarial attacks has posed threats to the applications in real-world scenarios, especially to some security-critical fields like autonomous driving and so on.

To counter adversarial attacks, many defense algorithms have been proposed, which can be broadly categorized into adversarial training \citep{madry2017towards,zhang2019theoretically,wu2020adversarial,dong2020adversarial,zhai2020macer}, input preprocessing \citep{dziugaite2016study, shen2017apegan, defensegan, HGD}, and adversarial detection \citep{xu2017featuresqueezing, ma2018characterizing, meng2017magnet, tian2021detecting}. Although these methods have achieved notable defense outcomes against known attacks, adversarial training suffers from high resource costs for training, and other methods inevitably exhibit vulnerabilities to unknown attacks or overfitting to specific attacks. Besides, they perform poorly in the face of advanced attack method \citep{athalye2018obfuscated}. That's to say, these approaches do not fundamentally address the problem in practical scenarios, some of which even lack theoretical guarantees of robustness and are susceptible to advanced attacks \citep{athalye2018obfuscated}. Recently, methods based on energy models \citep{hill2020stochastic} attempt to eliminate adversarial perturbations through Langevin sampling, yet their defensive accuracy remains modest. Some other existing works have demonstrated that adversarial sample defense with diffusion models has relatively high accuracy \citep{nie2022diffusion}, but they still fail to provide theoretical analysis. Recent research \citep{carlini2022certified} proposed to use diffusion models as adversarial defense modules and provided verifiable robustness under 2-norm constraints, without a detailed explanation of the adversarial defense mechanism of diffusion models. DENSEPURE \citep{xiao2022densepure} employs multiple rounds of reverse denoising steps to generate multiple samples, followed by a voting-based classification to determine the final result. However, this approach incurs substantial time consumption. Similarly, GuidedDiff \citep{nie2022diffusion} employs a similar denoising technique.

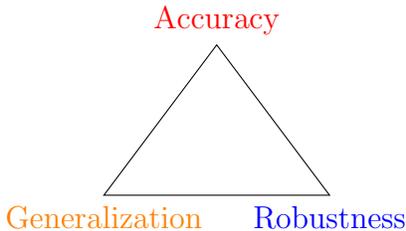
\begin{figure}[H]
    \centering
    \begin{tikzpicture}
        \coordinate (A) at (0,0);
        \coordinate (B) at (3,0);
        \coordinate (C) at (1.5,2);
        \draw (A) -- (B) -- (C) -- cycle;
        \fill[color=orange] (A) circle (0pt) node[below] {Generalization};
        \fill[color=blue] (B) circle (0pt) node[below] {Robustness};
        \fill[color=red] (C) circle (0pt) node[above] {Accuracy};
    \end{tikzpicture}
    \caption{Trade-off among accuracy, robustness, and generalization}
    \label{fig:trade-off}
\end{figure}

\begin{figure*}
    \centering
\includegraphics[width=0.7\textwidth,height=0.4\textwidth]{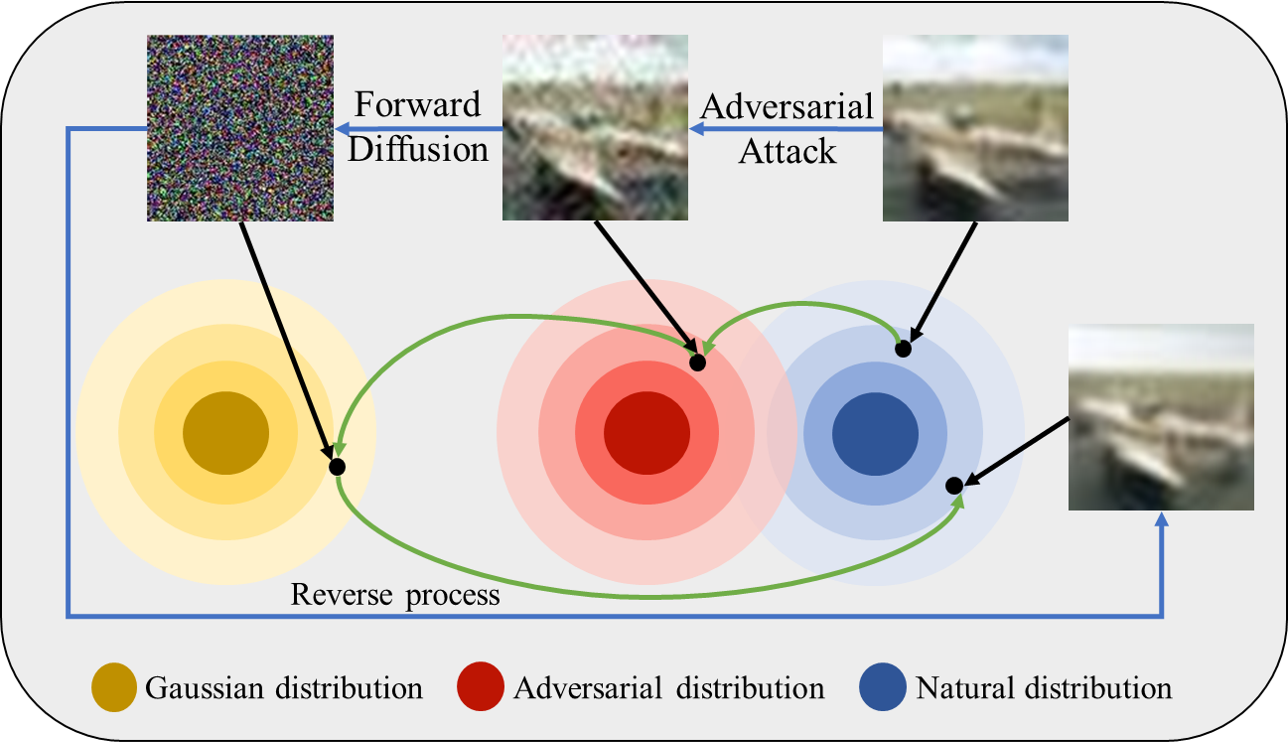}
    \caption{Diagram of the proposed method}
    \label{fig:overview}
\end{figure*}

On the whole, the existing defense methods do have demonstrated promising results on small-scale datasets, but they are not without limitations. Firstly, these defenses are prone to overfit the model or focus on specific attack methods, leading to poor generalization. Secondly, they lack a fundamental explanation of the adversarial sample generation process and lack a systematic framework. Lastly, there is an obvious trade-off among accuracy, robustness, and generalization with respect to these methods, as shown in Figure.\ref{fig:trade-off}.
To address these issues, this study takes a novel approach by examining the generation mechanism of adversarial examples and re-evaluating the relationship between adversarial and normal samples from the perspective of data distribution. Similar to the latest work\citep{ibrahim2022towards}, our approach extends the concept of out-of-distribution (OOD) generalization, unifying adversarial robustness and OOD generalization, and introducing a theoretical framework for defining and optimizing OOD generalization. Building upon this foundation, a defense strategy using distribution shift is proposed. Leveraging diffusion models trained on a large set of original training data, the method shifts the original sample distribution closer to a Gaussian distribution and then reverts it back to the original distribution (as illustrated in Figure \ref{fig:overview}). This paper achieves adversarial defense by harnessing this characteristic and incorporating model guidance while striking a balance between robustness and generalization, outperforming previous methods.

Specifically, the approach introduced in this paper aligns with input preprocessing, involving the use of plug-and-play preprocessing modules to mitigate the adversarial nature of samples. In recent years, diffusion models have gained traction across domains due to their remarkable image generation capabilities \citep{dhariwal2021diffusion}. Leveraging the exceptional image denoising and editing abilities of diffusion models \citep{lugmayr2022repaint}, researchers have ventured to utilize them as data preprocessing models to neutralize harmful samples. Wu et al. guided speech inputs through a diffusion model, eliminating sample adversariality via forward diffusion and reverse restoration processes \citep{wu2023defending}. In the image domain, scholars have proposed pretraining diffusion models on clean datasets and guided adversarial defense by evaluating the similarity between adversarial samples and reverse-reconstructed images \citep{wang2022guided,wu2022guided}. However, these methods predominantly focus on data and do not optimally harness the protected model's information, leading to diminished adaptability across various models. Addressing this limitation, this paper introduces the incorporation of the protected model's information during the reverse diffusion process, thereby enhancing the diffusion model's adaptability as a plug-and-play module for adversarial defense.

The main contributions of this paper are as follows:

\begin{itemize}
    \item We propose an adversarial defense method by unifying the concept of OOD generalization and adversarial robustness.
    \item Extensive empirical studies also validate that our method achieves a remarkable balance among accuracy, robustness, and generalization.
    \item We innovatively propose a new perspective on adversarial robustness, and theoretical analysis and experimental results demonstrate an intrinsic connection between adversarial robustness and OOD generalization.
\end{itemize}

The remainder of this paper is organized as follows. Section 2 reviews some related works about adversarial attack and defense. Section 3 introduces our insights into adversarial examples. Section 4 reports extensive experimental results that confirm the effectiveness of the proposed method.
Finally, we summarize this paper and discuss possible future work in
Section 5.

\section{Related work}

In this section, we briefly introduce key concepts for adversarial attack (Section \ref{section:2.1}) and defense (Section \ref{section:2.2}) and corresponding research that serve to better understand the rest of the paper.

\subsection{Adversarial attack} \label{section:2.1}

Adversarial attacks aim to incorporate carefully crafted subtle perturbations $\delta$ into original sample $x$, creating adversarial instance $x+\delta$ that make the target model $f_\theta(\cdot)$ misclassify the original sample. This objective can be mathematically expressed as follows:

\begin{equation}
\delta=\min _{\delta} \|\delta\|_p, f_\theta(x) \neq f_\theta\left(x+\delta\right)
\label{eq:aes}
\end{equation}
where $\|\cdot\|_p$ is the constraint on the adversarial perturbations. Generally, the $\infty$-norm ($p=\infty$) or the $2$-norm ($p=2$) is employed to bound the perturbation, ensuring the visual quality of the adversarial examples. Depending on the type of constraint, adversarial examples can be categorized into non-targeted and targeted attacks. Equation \ref{eq:aes} represents a non-targeted attack, which aims to induce misclassification in the model without specifying a particular target label. The attacker's level of knowledge about the victim model is classified into white-box, gray-box, and black-box attacks based on their knowledge of the model parameters, training data, and other information.

Despite the emergence of numerous adversarial defense techniques in recent years, Athalye pointed out the vulnerabilities in these methods and attacked them using Backward Pass Differentiable Approximation (BPDA) \citep{athalye2018obfuscated}. The findings indicated that the existing defense methods at that time were ineffective in countering adversarial attacks and merely offered certified security. Regarding defense approaches involving data preprocessing function $g$, like DefenseGAN \citep{defensegan}, if we assume that $g$ is non-differentiable, we can obtain the security classification model $h=g \circ f$, where $f$ represents the classifier or feature extractor. As $g$ is non-differentiable, obtaining gradients for the secure model's input becomes challenging, leading to the use of approximated gradients for attacks. However, as $f(x) \approx h(x)$ is required, we can approximate gradients $\nabla_{x}h(x)$ using $\nabla_{x}f(x)$. To validate the effectiveness of the proposed method with more accurate gradient estimation, in this paper, we conducted an adaptive attack using  $\nabla_{x}h(x)$ (here $g$ is differentiable) instead of $\nabla_{x}f(x)$.

\subsection{Adversarial defense}\label{section:2.2}

To address the aforementioned adversarial robustness challenge, researchers have put forth a range of adversarial defense methods. As a defender, it's intuitive to conduct adversarial detection to intercept malicious input. Feature Squeezing (FS) employs the vulnerability of adversarial examples for detection \citep{xu2017featuresqueezing}. The Local Intrinsic Dimensionality (LID) algorithm extracts multi-layer features from samples to distinguish between normal and adversarial samples \citep{ma2018characterizing}. MagNet detects adversarial examples by utilizing reconstruction errors and model output discrepancies of reconstructed images \citep{meng2017magnet}. However, these strategies are vulnerable to adaptive attacks, which can generate stealthy adversarial examples that bypass the detection methods.

Adversarial training is an effective and certified defense against adversarial attacks. Tradeoff-Inspired Adversarial Defense via Surrogate-Loss Minimization (TRADES) introduces consistency loss between adversarial and normal sample outputs to align their soft labels \citep{zhang2019theoretically}. Adversarial Weight Perturbation (AWP) enhances robustness by introducing weight perturbations as part of optimization, adjusting both inputs and weights \citep{wu2020adversarial}. Adversarial Distributional Training (ADT) attempts to train against adversarial examples using their distribution, seeking general adversarial training beyond isolated adversarial points \citep{dong2020adversarial}. Maximizing the Certified Radius (MACER) was a method agnostic training approach, elevating robustness by maximizing a model's verified robust radius \citep{zhai2020macer}. Nevertheless, adversarial training not only suffers from high training costs but also obtains adversarial robustness at the expense of accuracy.

Furthermore, methods based on input transformations utilize existing samples for reconstruction or preprocessing to diminish their adversarial impact. Classic image compression algorithm JPG can partly counter adversarial attacks \citep{dziugaite2016study}. DefenseGAN \citep{defensegan} reconstructs inputs using a generator to withstand adversarial attacks. High-Level Representation Guided Denoiser (HGD) preprocesses input samples using constrained high-dimensional features, offering improved generalization and defense capabilities \citep{HGD}. These types of methods, as mentioned before, are susceptible to adaptive attacks.

Overall, the existing defense methods fail to offer a practical choice. Their shortcomings can be summarized as follows: (1) high training cost, (2) trade-off between accuracy, robustness, and generalization, and (3) lack of certified robustness. Based on these deficiencies, we propose our distribution transfer-based defense method.

\section{Distribution transfer for adversarial defense}

\subsection{Understanding adversarial examples from a distribution perspective}
Existing studies have provided various explanations for the presence of adversarial examples (AEs) from different perspectives. Researchers like \citep{ilyas2019adversarialarefeatures} suggested that the existence of AEs is rooted in the non-robust features learned by deep learning models. These features significantly impact model decisions and consequently lead to misclassifications. Other research, such as that by \citep{fgsm}, interprets AEs in the context of linear explanations intrinsic to the model itself. Additionally, some investigations approach the topic from a high-dimensional perspective, indicating the weaker robustness of high-dimensional data against adversarial perturbations \citep{gilmer2018adversarial}.

These findings collectively highlight significant distributional disparities between adversarial and normal samples when viewed from a data distribution standpoint \citep{zhu2022toward,feinman2017detecting,xie2020adversarial}. Models demonstrate better performance within their training data distribution (in-distribution data), while their performance deteriorates considerably outside this distribution \citep{wald2021calibration,nagarajan2020understanding}. Zhu et al.'s research indicate a substantial divergence between the data distribution of adversarial and normal samples \citep{zhu2022toward}. This paper delves into the topic of AEs from the perspective of data distribution, aiming to provide a better understanding of their existence and characteristics as a foundation for subsequent adversarial defense research.

Based on the aforementioned research, a conclusion can be drawn that the existence of AEs arises from the vulnerability of models' generalization out of the training data distribution. In other words, models tend to exhibit poorer performance when dealing with data distributions far from the training data. To this end, this paper posits a hypothesis: \textit{adversarial robustness is essentially a corner case of models' OOD generalization, especially for complex high-dimensional data}. AEs are generated when the sampled data causes a more rapid change in the loss function than an in-distribution or OOD sample. During adversarial training, continuous sampling of AEs minimizes the influence of adversarial perturbations on model outputs, which can be formulated as:
\begin{equation}
\label{eq:eq1}
\begin{aligned}
& \mathbb{E}_{(x, y) \in D}\left[\frac{\|J(x,y; \theta)-J(\hat{x},y; \theta)\|_1}{\|x-\hat{x}\|_2}\right] \\
& \approx\mathbb{E}_{(x, y) \in D}\left[ \frac{(x-\hat{x})\cdot\nabla_x J(x, y;\theta)+o^n }{\|x-\hat{x}\|_2}\right]\\
& \approx \mathbb{E}_{(x, y) \in D}\left\|\nabla_x J(\hat{x}, y ; \theta)\right\|_2 \\
& \leq \mathbb{E}_{(x, y) \in D}\left\|\nabla_x J(T(x ; \theta), y ; \theta)\right\|_2 \\
& \leq \mathbb{E}_{(x, y) \in D}\left\|\nabla_x J(A(x;\theta), y ; \theta)\right\|_2
\end{aligned}
\end{equation}
Among these, $\hat{x}$ is sampled from in-distribution domain (small enough $\|\hat{x}-x\|_p \leq \epsilon$); $J(;\theta)$ represents the loss function of the model with parameters $\theta$, and $A(;\theta)$ and $T(;\theta)$ denote the adversarial attack and data augmentation methods for the model with parameters $\theta$. The above formula signifies that for each sample in this dataset $D$, when sampling is performed, the loss function exhibits greater changes corresponding to AEs than conventional data augmentation methods. Equation \ref{eq:eq1} also means that the model's performance on in-distribution samples, OOD samples, and AEs becomes worse in turn. That's to say, adversarial attacks generate AEs by sampling the corner case of the OOD samples. Therefore, we can unify the optimization of  OOD generalization and adversarial robustness, and recent works \citep{zhu2023improving,wang2023better,ibrahim2022towards} also verify our findings. Based on the analysis above, this paper optimizes OOD generalization:
\begin{equation}
\label{eq:eq2}
\min _\theta \mathbb{E}_{(x, y) \in S}[\ell(x, y ; \theta)]
\end{equation}
where $S=\left\{Q\left(X_Q, Y_Q\right) \mid \operatorname{Dist}(Q, D) \leq \rho\right\}$, $l(\cdot)$ represents the loss function and $Dist(\cdot)$ signifies a distribution metric, it can be inferred that the distance of from the training data distribution leads to a distribution. Consequently, employing data augmentation techniques \citep{schmidt2018adversarially} is a common practice to enhance a model's robustness beyond its training distribution. This augmentation is often executed through adversarial attacks $A(;\theta)$ and heuristic data augmentation algorithms $T(;\theta)$. Previous research has shown that adversarial training and conventional data augmentation methods contribute to reinforcing a model's OOD generalization for lower-dimensional data, such as in natural language processing tasks \citep{zhang2019you}. However, conventional augmentation techniques struggle to encompass the region occupied by AEs in scenarios involving high-dimensional data, like image-related challenges. As a result, the issue of adversarial robustness in high-dimensional data can be considered a much more specialized instance of OOD generalization. Similarly, even in the case of straightforward image data like CIFAR10, adversarial training proves effective in achieving impressive OOD generalization \citep{yi2021improved}. On the whole, from the perspective of distribution, we claim that there is no essential difference between OOD samples and AEs. As long as we can sample enough training data from $S$, the performance of OOD generalization can be guaranteed \citep{wang2023better}. However, the traditional adversarial training-based method is far from practical applications for its high hardware requirement (e.g., \citep{wang2023better} use NVIDIA A100 for adversarial training). Therefore, it's urgent for us to find a better way to optimize OOD generalization.

\subsection{Diffusion models}
Building upon the above analysis, we can infer that AEs also belong to the category of OOD samples, suggesting that optimization for OOD generalization can enhance a model's adversarial robustness. While a model may exhibit strong generalization within its in-distribution, the scarcity of training data constrains its performance when confronted with samples outside its training distribution. Similar to this notion, \citep{xie2020adversarial} utilized two batch normalization (BN) layers within a single model to separately train on adversarial and normal samples, resulting in a notable improvement in the model's OOD generalization. Inspired by these findings, we propose to transfer these OOD samples towards the in-distribution, thus boosting the model's performance on such OOD samples. Based on the aforementioned formula (3), we can transform the problem of OOD generalization into one of in-distribution generalization through distribution transfer.

This is accomplished by defining a distribution transfer function $K: X \rightarrow X$, which moves samples from an OOD to an in-distribution scenario ($S \rightarrow D$). This transformation yields the following problem: 

\begin{equation}
\label{eq:eq3}
\min _\theta \mathbb{E}_{(x, y) \in K(S)}{[\ell(x, y ; \theta)]}
\end{equation}

Given diffusion models' substantial distribution transfer capability, we employ diffusion models in this study to address the aforementioned problem. The process of diffusion models \citep{ho2020denoising} comprises forward diffusion and inverse recovery steps, effectively reverting a Gaussian distribution back to a specified target distribution. \citep{ho2020denoising} provide a general definition for this Markov process: 

\begin{equation}
\label{eq:eq4}
\begin{aligned}
& q\left(x_{1: T} \mid x_0\right)=\prod_{t=1}^T q\left(x_t \mid x_{t-1}\right) \\
& q\left(x_t \mid x_{t-1}\right)=\mathcal{N}\left(x_t ; \sqrt{1-\beta_t x_{t-1}, \beta_t I}\right)
\end{aligned}
\end{equation}
where $\left\{\beta_t \in(0,1)\right\}_{t=1}^T$ represents the variance parameter of Gaussian distributions in different iteration steps $t$, satisfying $\beta_1<\beta_2<\cdots<\beta_T$. It's worth noting that when $T \rightarrow \infty$, it corresponds to the standard Gaussian distribution: $x_T \sim \mathcal{N}(0, \mathbf{I})$. When $\alpha_t = 1 - \beta_t, \bar{\alpha_t} = \prod_{t=1}^{T}\alpha_t$, the following relationship holds: 

\begin{equation}
\label{eq:eq5}
q\left(x_t \mid x_0\right)=\mathcal{N}\left(x_t ; \sqrt{\bar{\alpha}_t} x_0,\left(1-\bar{\alpha}_t\right) \mathbf{I}\right)
\end{equation}

Subsequently, according to the process of inverse recovery, i.e., denoising a Gaussian distribution to align it with the target distribution $p(x_0)$: 

\begin{equation}
\label{eq:eq6}
    \begin{aligned}
& p_\theta\left(x_{0: T}\right)=p\left(x_T\right) \prod_{t=1}^T p_\theta\left(x_{t-1} \mid x_t\right) \\
& p_\theta\left(x_{t-1} \mid x_t\right)=\mathcal{N}\left(x_{t-1} ; \mu_\theta\left(x_t, t\right), \Sigma_\theta\left(x_t, t\right)\right)
\end{aligned}
\end{equation}
where $\mu_{\theta}(x_t,t),\sum_{\theta}(x_t,t)$ and are parameterized models, and a time-dependent constant.

\subsection{Adversarial defense based on distribution transfer}

Drawing from the preceding research, this study presents a novel approach to adversarial defense by employing diffusion models for distribution transfer. Notably, similar to prior work, preserving semantic consistency during sample generation, while simultaneously transitioning AEs from an OOD context to an in-distribution state, stands as a pivotal requirement to achieve effective adversarial defense while minimizing perturbations to the model's baseline performance\citep{wu2023defending}. The proposed adversarial defense process in this study is defined as follows \citep{dhariwal2021diffusion}: 

\begin{equation}
\label{eq:eq7}
\begin{aligned}
& p_\theta\left(x_{t-1} \mid x_t, x_0^{\prime}\right)=C_1 p_\theta\left(x_{t-1} \mid x_t\right) p\left(x^{\prime} \mid x_t\right) \\
& =C_1 \mathcal{N}\left(\mu+\Sigma \nabla_{x_t} \log p\left(x^{\prime} \mid x_t\right), \Sigma\right)
\end{aligned}
\end{equation}
where $x^{\prime}$ denotes the AEs fed into the model. In alignment with related approaches \citep{wang2022guided,wu2022guided}, a heuristic methodology is adopted to steer the process of distribution transfer. However, a notable divergence lies in the recognition of the distribution incongruity between adversarial and original samples, coupled with the unavailability of target distribution insights. Consequently, relying solely on prior knowledge derived from training data proves insufficient to achieve the desired outcome. Yet, drawing insights from prior adversarial defense endeavors \citep{xu2017featuresqueezing,dziugaite2016study} which underscore the vulnerabilities of AEs, distribution transfer can be guided by exploiting this intrinsic vulnerability: 

\begin{equation}
\label{eq:eq8}
p\left(x_0^{\prime} \mid x_t\right)=C_2 \exp \left(-s \cdot \mathcal{D}\left(x_t^{\prime}, x_t\right)\right)
\end{equation}
where $\mathcal{D}(\cdot)$ represents a distance metric between two samples while signifies a scaling factor. Accordingly, the previously mentioned equation (4.5) can be expressed as follows: 

\begin{equation}
\label{eq:eq9}
p_\theta\left(x_{t-1} \mid x_t, x_0^{\prime}\right)=C_1 \mathcal{N}\left(\mu-s \Sigma \nabla_{x_t} \mathcal{D}\left(x_t^{\prime}, x_t\right), \Sigma\right)
\end{equation}

Traditional methods predominantly rely on Mean Squared Error (MSE) \citep{wu2022guided}, Structural Similarity Index (SSIM) (Wang et al., 2022), or adopt an unguided approach \citep{cohen2019certified,xiao2022densepure} for sample inversion. However, these methods rely solely on the diffusion model, lacking auxiliary information that encapsulates target distribution characteristics. To surmount this limitation, we introduce a distance metric and reference model output, incorporating both source and target distribution information to guide the process of reverse sample generation by the diffusion model: 

\begin{equation}
\label{eq:eq10}
\mathcal{D}\left(x_t^{\prime}, x_t\right)=\left\|f\left(x_t^{\prime} ; \theta\right)-f\left(x_t ; \theta\right)\right\|_2+\varphi \cdot \operatorname{SSIM}\left(x_0^{\prime}, x_t\right)
\end{equation}
where $\varphi$ denotes a coordination factor; the first part introduces distribution-related information, while the latter ensures controlled sample generation.

\subsection{Robustness analysis}

The random smoothing method \citep{cohen2019certified} is often employed to assess a model's $l_2$ robustness, achieved by introducing random noise to obtain a smoothed classifier: 

\begin{equation}
\label{eq:eq11}
g(x)=\arg \max _c \mathbb{P}_{\delta \in \mathcal{N}\left(0, \sigma^2 I\right)}\left(f_\theta(x+\delta)=c\right)
\end{equation}
where $\sigma$ denotes the extent of perturbation. According to this definition, the robust radius of the smoothed classifier can be expressed as: 

\begin{equation}
\label{eq:eq12}
R=\frac{\sigma}{2}\left(\Phi^{-1}\left(p_A\right)-\Phi^{-1}\left(p_B\right)\right)
\end{equation}
where $\Phi(\cdot)$ signifies the standard Gaussian cumulative distribution function, and $p_A$ and $p_B$ correspond to the probabilities of the highest and second-highest classes, respectively. This approach was also utilized in the experiments to examine the proposed method.

Moreover, a demonstration of the $l_2$ robustness of the proposed method is provided. Following the approach above-mentioned, the diffusion model $H:X \rightarrow X$ is introduced, effectively substituting the original classifier with $f(H(\cdot))$, and ensuring that the samples $x^{\prime}$ subjected to diffusion adhere to the following criteria: 

\begin{equation}
\label{eq:eq13}
\left\|x^{\prime}-x\right\| \leq \delta+\sqrt{e^{2 \gamma\left(t^*\right)}-1} \cdot C_\alpha+\gamma\left(t^*\right) \cdot C_s
\end{equation}
where $\delta$ represents the adversarial perturbation, and $C_{\alpha}$ and $C_{s}$ are constants, a proof of which can be referred to \citep{nie2022diffusion}. Substituting into equation (13), we obtain: 

\begin{equation}
\label{eq:eq14}
R=\frac{\left(\delta+\sqrt{e^{2 \gamma\left(t^*\right)}-1} \cdot C_\alpha+\gamma\left(t^*\right) \cdot C_S\right)}{2} \times \left(\Phi^{-1}\left(p_A\right)-\Phi^{-1}\left(p_B\right)\right)
\end{equation}

This directly yields the robust radius calculated by our method, thereby ensuring the stability of the approach proposed in this study.

\begin{table*}[]
\centering
\caption{Accuracy($\%$) of different adversarial defense methods on CIFAR10 dataset}
\label{tab:tab1}
\setlength{\tabcolsep}{6pt}
\renewcommand{\arraystretch}{1.2}
\begin{tabular}{ccccccccc}
\toprule
\multirow{3}{*}{} & \multirow{2}{*}{\makecell{Extra \\Data}} & \multirow{2}{*}{\makecell{Extra \\Model}} & \multicolumn{3}{c}{ResNet18} & \multicolumn{3}{c}{WRN28-10} \\
 &  &  & Clean & $l_{\inf}^{adv}$ & $l_2^{adv}$ & Clean & $l_{\inf}^{adv}$  & $l_2^{adv}$ \\
\midrule
AT & \ding{52} & \ding{56} & 74.4 & 65.8 & 79.9 & 76.2 & 67.7 & 76.8 \\
AT-ALP & \ding{52} & \ding{56} & 72.1 & 60.3 & 71.3 & 80.7 & \textbf{70.3} & \textbf{80.5} \\
TRADES & \ding{52} & \ding{56} & 80.1 & 71.1 & \textbf{84.5} & 77.8 & 59.8 & 74.6 \\
TRADE-AWP & \ding{52} & \ding{56} & 68.1 & 63.6 & 70.6 & - & - & - \\
MACER* & \ding{52} & \ding{56} & 82.3 & 43.3 & 43.6 & - & - & - \\
\midrule
EBM-Defense & \ding{56} & \ding{52} & - & - & - & 75.8 & 51.2 & 50.1 \\
DensePure & \ding{56} & \ding{52} & 74.2 & 35.6 & 44.5 & 61.9 & 39.5 & 39.3 \\
Ours & \ding{56} & \ding{52} & \textbf{86.7} & \textbf{78.1} & 74.3 & \textbf{84.2} & 66.5 & 66.3 \\
\bottomrule
\end{tabular}
\end{table*}

\section{Experiments}

\subsection{Experimental setups}
The adversarial defense method proposed in this paper is implemented based on the PyTorch deep learning framework. The specific software environment includes Windows 10 operating system, Python 3.8 programming language, and PyTorch version 1.11. The hardware setup comprises an Intel I9 9900K CPU, 32 GB of RAM, and an NVIDIA RTX 2080Ti GPU.

\subsection{ Datasets $\&$ models}
To validate the performance of our proposed method, we employed the following models: ResNet18, ResNet50 \citep{he2016deep}, MobileNetv2 \citep{sandler2018mobilenetv2}, ShuffleNetv2 \citep{ma2018shufflenet}, WideResNet28-10 \citep{zagoruyko2016wide}, and Inception-ResNetv2 \citep{szegedy2017inception}. These models were experimented on the CIFAR10 and a subset of the ImageNet dataset, referred to as ImageNet30.

For evaluating adversarial robustness, we employed the PGD attack method as well as the self-adaptive attack algorithm as mentioned in section \ref{section:2.1}. We generated adversarial samples using two different optimization losses and tested the adversarial defense performance of our proposed method.

\subsection{Evaluation metrics}
The proposed method was evaluated from multiple aspects, including model accuracy, adversarial robustness, and generalization capability. Model accuracy was assessed by the accuracy of the model on the clean test set samples. Adversarial robustness was evaluated using the accuracy of adversarial samples generated from the clean test set under 2-norm (128/255) and infinity-norm (8/255) perturbations. The CIFAR10-C dataset \citep{hendrycks2018benchmarking} was employed to measure generalization capability. This dataset consists of samples generated through 15 types of transformations, such as noise, blur, and weather, each with five levels of severity, resulting in a total of 75 scenarios.

Furthermore, for evaluating the quality of the reverse reconstructed images, objective evaluation metrics, including the Structural Similarity Index (SSIM) and Peak Signal-to-Noise Ratio (PSNR), as well as the Learned Perceptual Image Patch Similarity (LPIPS) \citep{zhang2018unreasonable} were utilized. The first two metrics are objective measures, while the third is based on pre-trained models' feature similarity scores.

\subsection{Baseline method}
This paper primarily contrasts existing adversarial defense methods based on adversarial training and diffusion models. Specifically, the covered methods include traditional adversarial training method AT\citep{madry2017towards}, adversarial sample defense based on logits pair AT-ALP\citep{kannan2018adversarial}, TRADES \citep{zhang2019theoretically}, adversarial defense with weight perturbation TRADES-AWP \citep{wu2020adversarial}, diffusion-based method DENSEPURE \citep{xiao2022densepure}, as well as energy-based model approaches MACER \citep{zhai2020macer} and EBM (Energy based model) defense \citep{hill2020stochastic}.

\subsection{Results and discussion}

\subsubsection{Performance of adversarial defense}
In this study, adversarial samples generated by PGD attacks (40 iterations) under two norm constraints ($l_{\inf}$: 128/255, $l_2$: 8/255) were evaluated on the CIFAR10 test set. The diffusion model underwent forward iterations for a total of 470 times. The robustness of different comparison methods was assessed. Additionally, a comparison was made regarding whether different methods utilized additional data (including adversarial training) or required extra models. The experimental results are presented in Table \ref{tab:tab1}. Notably, the proposed approach exhibited superior performance on clean data, significantly outperforming adversarial training and other data preprocessing methods. Remarkably, leveraging only a pre-trained diffusion model enabled ResNet18 to achieve an accuracy of 86.7\%. For infinite-norm adversarial samples, the proposed method outperformed adversarial training by 7\%. It's worth noting that the defense performance is somewhat limited for 2-norm adversarial samples. This could possibly be attributed to the robustness introduced by the Gaussian noise added during the diffusion model's processing of 2-norm adversarial perturbations. Nevertheless, even under these circumstances, the strategy proposed in this paper still demonstrated notable defense performance.

\subsubsection{The influence of perturbation budgets}
\begin{figure}
    \centering
\includegraphics[width=0.6\textwidth,height=0.4\textwidth]{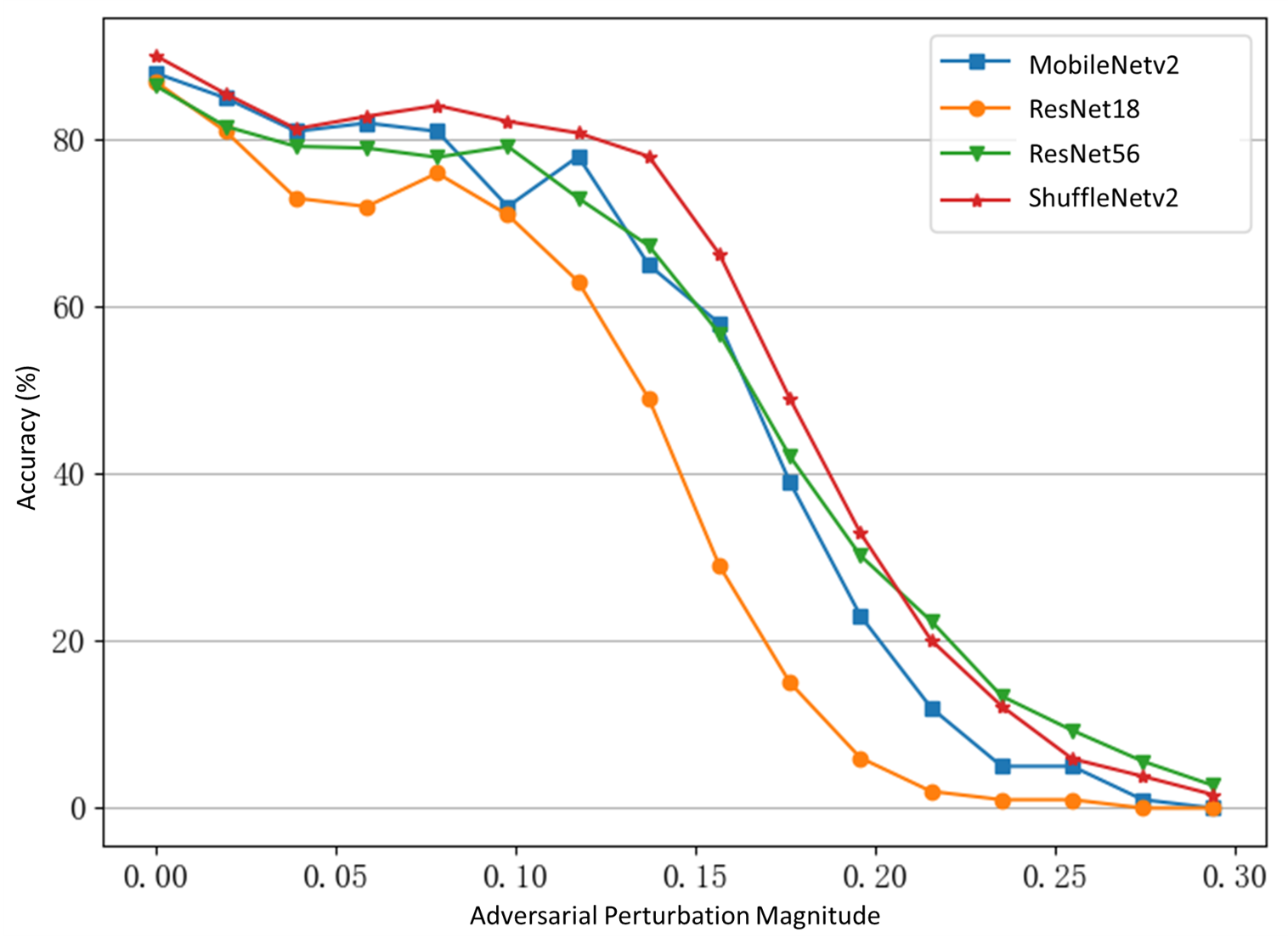}
    \caption{Impact of adversarial perturbation magnitude on accuracy}
    \label{fig:pert_budget}
\end{figure}
To investigate the enhancement of diffusion models on different model robustness under varying perturbation magnitudes, this study focused on the CIFAR10 dataset. Using PGD attacks ($l_2$: 8/255, 40 iterations) and setting the diffusion model's forward iterations to 470, the experiments assessed adversarial perturbations ranging from 0 to 80/255. Each evaluation involved 1000 samples, and the results are illustrated in Figure \ref{fig:pert_budget}. It indicates that diffusion models exhibit commendable defense performance when perturbation magnitude is confined within 0.1. However, as the perturbation magnitude increases, the effectiveness of adversarial defense deteriorates rapidly. This phenomenon stems from significant distributional differences between adversarial samples and the original distribution.
Consequently, the subsequent investigation will explore the impact of diffusion iteration count on adversarial defense capability. With increased diffusion iterations, adversarial samples are anticipated to transition from an adversarial or OOD distribution to a Gaussian distribution. This transition would enable a better mapping within the denoising process, facilitating accurate recognition by the model.

\subsubsection{The influence of forward steps}
\begin{figure}
    \centering
\includegraphics[width=0.6\textwidth,height=0.4\textwidth]{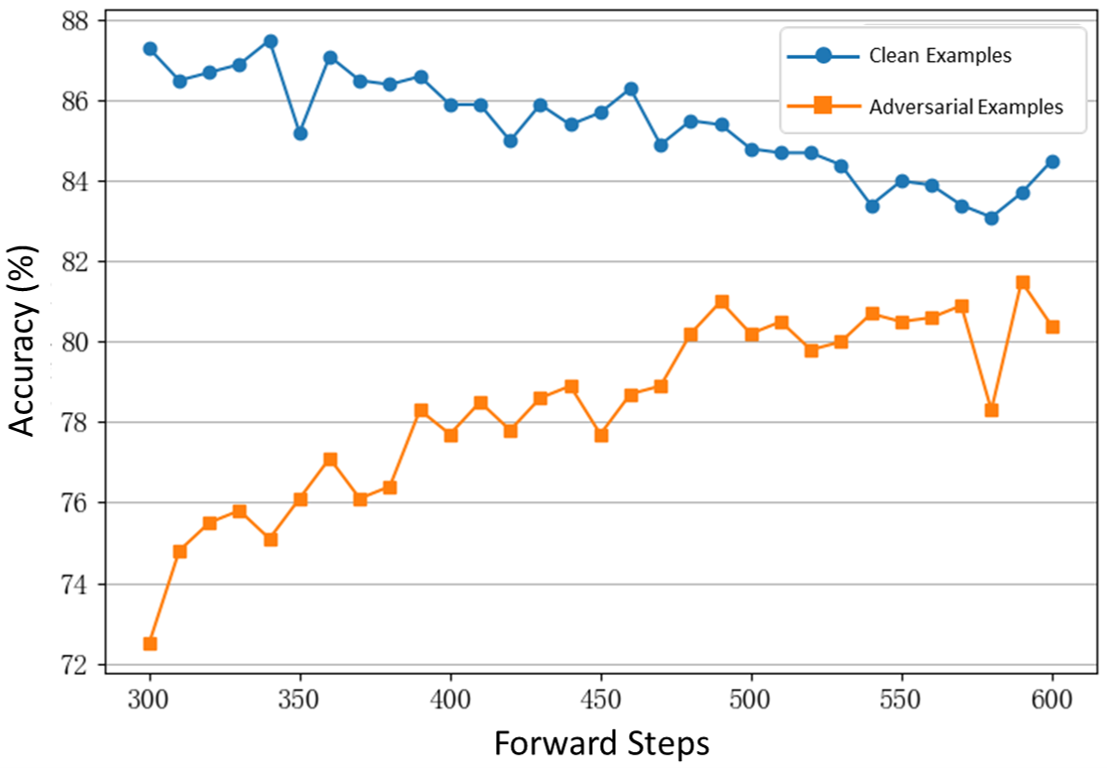}
    \caption{Impact of forward process steps on accuracy}
    \label{fig:forward_steps}
\end{figure}
\begin{figure}
    \centering
\includegraphics[width=0.5\textwidth,height=0.4\textwidth]{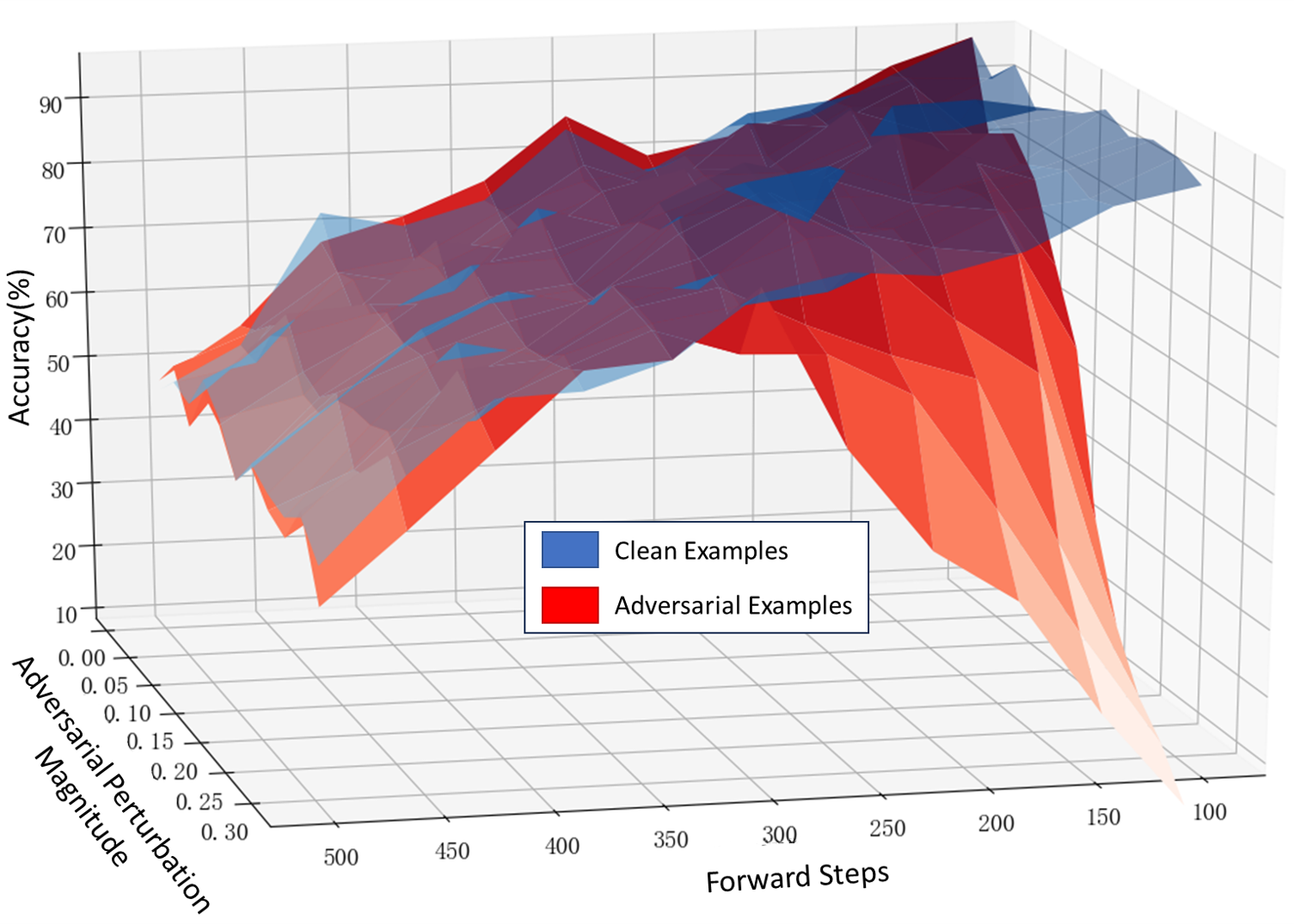}
    \caption{Impact of perturbation magnitude and forward process steps on accuracy}
    \label{fig:3d_vis}
\end{figure}

In the practical implementation of forward diffusion, it's essential to impose constraints on the diffusion iteration count. If the iteration count is high, the result of inverse denoising may deviate significantly from the original image, thus compromising consistency. Conversely, a low iteration count might not effectively transition the sample from the distribution of adversarial examples to that of normal examples. Considering these factors, experimentation is required to explore the impact of iteration count on the experimental outcomes. Employing MobileNetv2 as the victim model and the PGD attack ($l_2$: 8/255, 40 iterations), a study was conducted on the CIFAR10 dataset, as depicted in Figure \ref{fig:forward_steps}. The observations suggest that within a certain range, as the diffusion steps increase, the model's defense against adversarial samples gradually improves, whereas its performance on clean samples deteriorates. This trend aligns with the earlier conjecture, i.e., an increase in forward diffusion steps may lead to a drop in accuracy for clean samples and an increase in accuracy for adversarial samples. The same experiment was carried out on the ImageNet30 dataset (using ShuffleNetv2), and the test results are illustrated in Figure \ref{fig:3d_vis}. The findings are consistent with those on the CIFAR10 dataset, indicating that for original samples, fewer diffusion iterations result in higher accuracy, while for adversarial samples, there exists an optimal iteration count to maintain accuracy within a certain range.

\subsubsection{Adaptive attack}

\begin{table}[]
\centering
\caption{Performance ($\%$) on adaptive attack}
\label{tab:tab2}
\setlength{\tabcolsep}{12pt}
\renewcommand{\arraystretch}{1.2}
\begin{tabular}{ccccc}
\toprule
& \multicolumn{2}{c}{CIFAR10} & \multicolumn{2}{c}{ImageNet30} \\
& Clean & $l_{\inf}^{adv}$ &  Clean & $l_{\inf}^{adv}$ \\
\midrule
MobileNetv2 & 85.4 & 80.9 & 75.5 & 70 \\
ShuffleNetv2 & 90.4 & 83.9 & 78.5 & 78.0 \\
ResNet56 & 87.4 & 81.2 & - & - \\
ResNet18 & 86.2 & 82.5 & 68.0 & 69.5 \\
WRN28-10 & 83 & 72.4 & - & - \\
InceptionRNv2 & - & - & 66.0 & 61.0 \\
\bottomrule
\end{tabular}
\end{table}

Faced with conventional attack methods, although the diffusion-based adversarial distribution migration technique achieves considerable defense against adversarial samples, such data preprocessing methods fundamentally exploit the inconsistency in the OOD generalization capabilities between diffusion and classification models, leading to spurious robustness. While a diffusion model trained on clean datasets can effectively shift the data distribution for normal samples, as previously mentioned, the distribution of adversarial samples differs. Considering this aspect, if the generalization capabilities of the diffusion model and classification models' generalization capabilities become more aligned, this defense strategy could potentially lose effectiveness. Subsequently, an adaptive attack will be applied to both the classifier and the diffusion model to test this conjecture. As evident from Table \ref{tab:tab2}, even when an adaptive attack is targeted at the defense module, it has minimal impact on the original performance, further reinforcing the validity of the approach proposed in this paper.

\begin{table*}[]
\centering
\caption{Accuracy ($\%$) of different adversarial defense methods on CIFAR10-C dataset}
\label{tab:tab3}
\setlength{\tabcolsep}{4pt}
\renewcommand{\arraystretch}{1.2}
\begin{threeparttable}
\begin{tabular}{ccccccccccc}
\toprule
\multirow{2}{*}{} & 
\multicolumn{2}{c}{\ding{172}} & 
\multicolumn{2}{c}{\ding{173}} & 
\multicolumn{2}{c}{\ding{174}} & 
\multicolumn{2}{c}{\ding{175}} & 
\multicolumn{2}{c}{\ding{176}} \\ 
\cline{2-11} 
 & w & w/o & w & w/o & w & w/o & w & w/o & w & w/o \\ 
\midrule
MobileNetv2 & 86.7 & 57.4 & 76.8 & 50.8 & 82.8 & 71.4 & 82.3 & 78.6 & 86.0 & 73.0 \\
ShuffleNetv2 & \textbf{90.3} & 72.1 & \textbf{81.6} & 53.4 & \textbf{88.1} & \textbf{82.4} & \textbf{89.1} & 87.4 & \textbf{92.1} & 82.4 \\
ResNet56 & 86.7 & 66.4 & 76.2 & 41.6 & 84.3 & 76.8 & 85.3 & 76.3 & 87.0 & 75.5 \\
ResNet18 & 88.5 & 65.6 & 76.1 & 42.1 & 86.3 & 77.9 & 84.7 & 79.7 & 87.5 & 76.7 \\
WRN28-10 & 83.8 & 65.1 & 70.9 & 36.3 & 82.8 & 74.6 & 82.9 & 78.0 & 84.4 & 76.0 \\ 
\midrule
AT & 
\multicolumn{2}{c}{81.2} & 
\multicolumn{2}{c}{76.5} & 
\multicolumn{2}{c}{79.6} & 
\multicolumn{2}{c}{80.8} & 
\multicolumn{2}{c}{81.2} \\
AT-ALP & 
\multicolumn{2}{c}{73.1} & 
\multicolumn{2}{c}{68.4} & 
\multicolumn{2}{c}{73.1} & 
\multicolumn{2}{c}{72.5} & 
\multicolumn{2}{c}{73.8} \\
TRADES & 
\multicolumn{2}{c}{84.3} & 
\multicolumn{2}{c}{79.1} & 
\multicolumn{2}{c}{82.3} & 
\multicolumn{2}{c}{84.0} & 
\multicolumn{2}{c}{85.6} \\
TRADES-AWP & 
\multicolumn{2}{c}{73.5} & 
\multicolumn{2}{c}{71.0} & 
\multicolumn{2}{c}{73.1} & 
\multicolumn{2}{c}{73.5} & 
\multicolumn{2}{c}{73.3} \\
MACER* & 
\multicolumn{2}{c}{78.2} & 
\multicolumn{2}{c}{74.6} & 
\multicolumn{2}{c}{77.8} & 
\multicolumn{2}{c}{77.8} & 
\multicolumn{2}{c}{78.0} \\
EBM-Defense & 
\multicolumn{2}{c}{72.9} & 
\multicolumn{2}{c}{48.5} & 
\multicolumn{2}{c}{58.4} & 
\multicolumn{2}{c}{78.7} & 
\multicolumn{2}{c}{80.3} \\ 
\bottomrule
\end{tabular}
\begin{tablenotes}
\footnotesize
\item Note that \ding{172}\--\ding{176} stand for Gaussian Noise, Glass Blur, Impulse Noise, Jpeg Compression and Shot Noise respectively.
\end{tablenotes}
\end{threeparttable}
\end{table*}

\subsubsection{The performance of out-of-distribution generalization}

To validate the preceding conjecture of this paper that adversarial samples are a special case of OOD generalization, and to elevate the model's performance on conventional OOD data by implementing adversarial sample defense, experiments were conducted on the CIFAR10-C dataset. In addition to adversarial defense, this paper further tested the OOD (with substantial distribution differences) dataset CIFAR10-C, and the results are presented in Table \ref{tab:tab3}. Data samples subjected to various transformations exhibit distinct distribution disparities compared to clean samples. Despite this, the experimental outcomes strongly indicate that the defense method based on distribution migration showcases defense efficacy against adversarial samples and significantly restores the general OOD data closer to the original distribution. As a result, it enhances the model's accuracy on OOD data. Compared to scenarios without a diffusion model, the approach achieved an average enhancement of 12.7\% in performance. Notably, although many other adversarial defense methods (ResNet18 model) generally fall short of the method proposed in this paper, they still outperform cases where diffusion model-based data preprocessing is absent. This signifies that enhancing a model's adversarial robustness can bolster the model's generalization performance.

% \begin{table}[htbp]
% \centering
% \caption{Objective image quality assessment metrics of generated images with diffusion model}
% \label{tab:tab4}
% \setlength{\tabcolsep}{12pt}
% \renewcommand{\arraystretch}{1.2}
% \begin{tabular}{cccc}
% \toprule
% & SSIM$\downarrow$ & PSNR$\uparrow$ & LPIPS$\downarrow$ \\
% \midrule
% Adversarial Examples & 0.044 & 32.556 & 0.090 \\
% Denoised Examples & 0.291 & 9.823 & 0.456 \\
% \bottomrule
% \end{tabular}
% \end{table}

\subsubsection{The quality of the generated data}
\begin{figure*}
\centering

\begin{minipage}{0.48\linewidth}
\centering
\includegraphics[width=1.1\textwidth,height=0.65\textwidth]{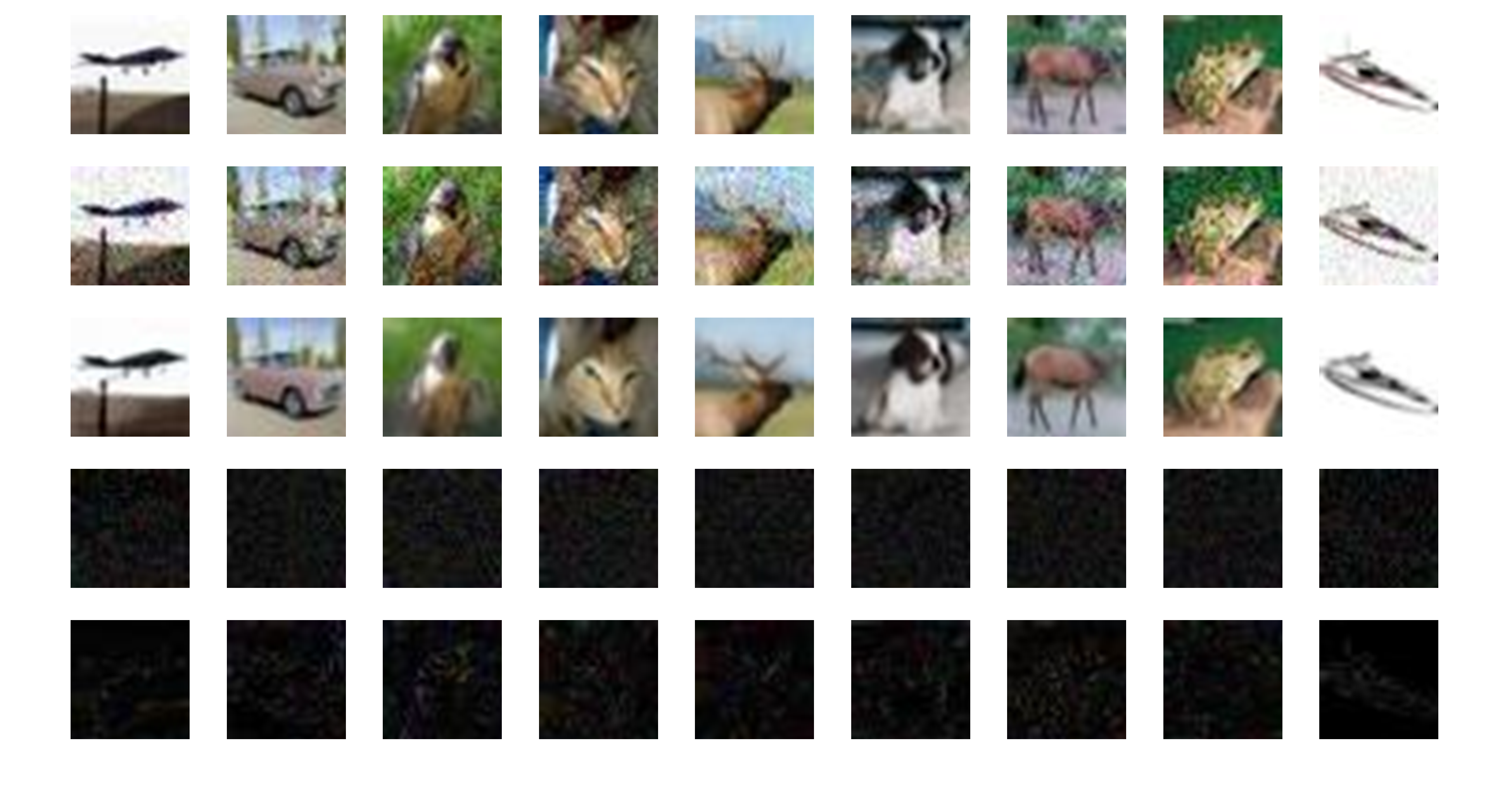}
\subcaption{CIFAR10}
\label{fig:CIFAR10_vis}
\end{minipage}
\begin{minipage}{0.48\linewidth}
\centering
\includegraphics[width=0.9\textwidth,height=0.65\textwidth]{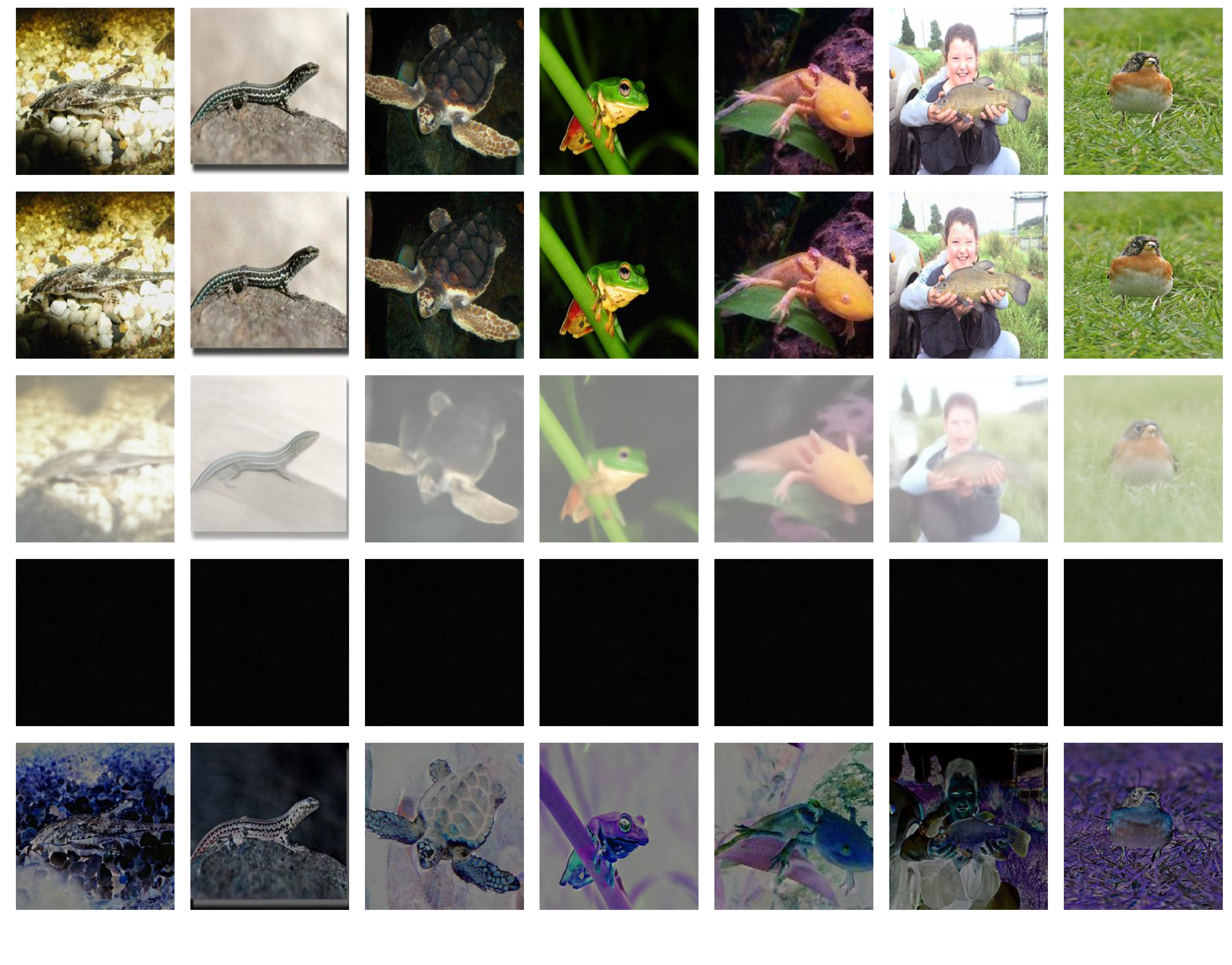}
\subcaption{ImageNet30}
\label{fig:ImageNet30_vis}
\end{minipage}

\caption{Visualization of clean images, adversarial samples, generated images, adversarial perturbations, and generated perturbations}
\label{fig:image_vis}
\end{figure*}

To ensure the quality of generated samples, this paper visualized the noise of original samples, adversarial samples, adversarial samples denoised by the diffusion model, adversarial perturbations, and denoised samples for both CIFAR10 dataset (Figure \ref{fig:CIFAR10_vis}) and ImageNet30 dataset (Figure \ref{fig:ImageNet30_vis}). Observing Figure \ref{fig:image_vis}, it's evident that the noise patterns of adversarial perturbations and denoised samples diverge. This divergence signifies that the principle underlying adversarial defense using the diffusion model and past denoising-based adversarial defense methods differ. Consequently, the proposed method does not risk overfitting to a specific adversarial attack, showcasing a solid capability for transferable adversarial defense. Although significant changes in the samples extracted from ImageNet30 perturbations were observed after preprocessing, the essential semantic information was well preserved.

% In addition to subjective image quality evaluation, this paper also provided objective image evaluation metrics (averaged over 200 data) between original, adversarial, and transferred adversarial samples \ref{tab:tab4}. The results indicate that due to the constraints on adversarial perturbations during adversarial sample generation, both SSIM and PSNR between adversarial samples and original samples significantly surpass those between denoised samples and original samples. However, concerning the learned perceptual image patch similarity (LPIPS), denoised samples outperform adversarial samples by a considerable margin. This implies that the efficacy of adversarial perturbations is diminished after being processed by the diffusion model, effectively achieving robust defense.

\section{Conclusion}

This paper centers its research on adversarial defense, aiming to enhance the adversarial robustness of models and address issues like insufficient generalization, high defense costs, and weak interpretability associated with existing methods. Taking a perspective grounded in data distribution, this study approaches the challenges of both adversarial robustness and out-of-distribution generalization of deep learning models from a unified angle, establishing a theoretical framework to bolster the ability for out-of-distribution generalization.

Built upon this theoretical framework, this paper introduces a strategy for adversarial defense through distribution migration. Leveraging a pre-trained diffusion model as its core, which is trained on a substantial amount of original data, the model spreads the original samples away from their distribution and then reconvenes them back to the original distribution. This paper exploits this inherent characteristic to guide the denoising process in a reverse manner, achieving adversarial defense. Experimental results underscore that utilizing the diffusion model for defense significantly enhances the model's generalization ability, effectively combining robustness and generalization. The method demonstrates robust defense performance even when the attacker is aware of the defense approach, surpassing the performance of existing methods.

By employing distribution migration for adversarial defense, this paper has addressed several issues present in prior works. The challenge of achieving defense with smaller costs in more complex scenarios constitutes a potential avenue for future research.

\section{Declaration of competing interest}
The authors declare that they have no known competing financial interests or personal relationships that could have appeared to influence the work reported in this paper.

\section{Acknowledgments}
Diqun Yan is partially supported by the National Natural Science Foundation of China (62171244, 61901237) and Ningbo Science and Technology Innovation Project (2022Z074, 2022Z075)



\bibliographystyle{elsarticle-harv}
\bibliography{reference.bib}

\end{document}